\documentclass[conference]{IEEEtran}
\usepackage[dvips]{graphicx}
\usepackage{amsmath}
\usepackage{amsthm}
\usepackage[flushleft]{threeparttable}
\usepackage{array}
\usepackage{booktabs}
\usepackage{bbm}
\newcommand{\PreserveBackslash}[1]{\let\temp=\\#1\let\\=\temp}
\newcolumntype{C}[1]{>{\PreserveBackslash\centering}p{#1}}
\newcolumntype{R}[1]{>{\PreserveBackslash\raggedleft}p{#1}}
\newcolumntype{L}[1]{>{\PreserveBackslash\raggedright}p{#1}}

\usepackage{bm}
\usepackage{graphicx,subfigure}
\usepackage{epstopdf}
\usepackage{algorithm}
\usepackage{cite}
\usepackage{amsmath}
\usepackage{amssymb}
\usepackage{psfrag}
\usepackage{algpseudocode}
\usepackage{empheq}
\usepackage{latexsym, amsmath, subfigure, color, amsfonts, amssymb,graphicx}
\usepackage{wrapfig}
\usepackage{enumitem}
\usepackage{caption}

\usepackage{bbm, dsfont}
\newtheorem{Theorem}{Theorem}

\newtheorem{Definition}{Definition}

\newtheorem{proposition}{Proposition}
\IEEEoverridecommandlockouts
\usepackage{tabularx}

\usepackage[T1]{fontenc}
\usepackage{etoolbox}

\makeatletter
\def\hlinewd#1{%
\noalign{\ifnum0=`}\fi\hrule \@height #1 %
\futurelet\reserved@a\@xhline}
\patchcmd{\maketitle}{\@fnsymbol}{\@alph}{}{}  
\makeatother
\def\hlinewd#1{%
\noalign{\ifnum0=`}\fi\hrule \@height #1 %
\futurelet\reserved@a\@xhline}
\patchcmd{\maketitle}{\@fnsymbol}{\@alph}{}{}  
\makeatother

\title{Centralized Caching and Delivery of Correlated Contents over a Gaussian Broadcast Channel}

\author{
\thanks{This work has been supported by NSF  Grant \#1619129, and by the European Research Council (ERC) Staring Grant BEACON (grant agreement No. 725731).}
    \IEEEauthorblockN{Qianqian Yang\IEEEauthorrefmark{1}, Parisa Hassanzadeh\IEEEauthorrefmark{2}, Deniz G\"und\"uz\IEEEauthorrefmark{1}, Elza Erkip\IEEEauthorrefmark{2}}
    \IEEEauthorblockA{\IEEEauthorrefmark{1}Electrical and Electronic Engineering Department, Imperial College London, London SW7 2AZ, U.K.
    \\\{q.yang14, d.gunduz\}@imperial.ac.uk}
    \IEEEauthorblockA{\IEEEauthorrefmark{2} 
     Electrical and Computer Engineering Department, New York University, Brooklyn, NY
    \\\{ph990, elza\}@nyu.edu}
}

\date{}

\begin{document}

\maketitle
\begin{abstract}
Content delivery in a multi-user cache-aided broadcast network is studied, where a server holding a database of correlated contents communicates with the users over a Gaussian broadcast channel (BC). The minimum transmission power required to satisfy all possible demand combinations is studied, when the users are equipped with caches of equal size. A lower bound on the required transmit power is derived, assuming uncoded cache placement, as a function of the cache capacity. A centralized joint cache and channel coding scheme is proposed, which not only utilizes the user's local caches, but also exploits the correlation among the contents in the database. This scheme provides an upper bound on the minimum required transmit power for a given cache capacity. Our results indicate that exploiting the correlations among the contents in a cache-aided Gaussian BC can provide significant energy savings.    
\end{abstract}

\section{Introduction}\label{intro}
Thanks to the decreasing cost and increasing capacity of memory available at mobile devices, \textit{proactive caching} has been considered as a low-cost and effective solution to today's exponentially growing mobile data traffic\cite{samuel2017,MaddahAliCentralized, MohammadQianDenizITW}.
Proactively storing popular contents in cache memories distributed across the network during off-peak traffic periods can greatly reduce both the network congestion and the latency during peak traffic hours. \textit{Coded caching} \cite{MaddahAliCentralized} further exploits the broadcasting nature of wireless delivery and the contents available in users' local cache memories to create multicasting opportunities, even when the users request distinct files, which further boosts the benefits of caching. The significant gains of coded caching over traditional uncoded caching schemes have inspired numerous studies, among which \cite{ hassanzadeh2016correlation,qiandeniz2018icc, hassanzadeh2017broadcast, hassanzadeh2017rate, hassanzadeh2016cache,hassanzadehJSAC,yu2017exact, MohammadDenizerasureTCom, amiri2017gaussian, shirin2016broadcastit, shirin2017broadcastit, Shengerasure2016} are most related to this paper.   

Delivering correlated contents over an error-free shared cache-aided link is considered in \cite{hassanzadeh2016correlation, hassanzadeh2017broadcast, hassanzadeh2017rate,hassanzadehJSAC,qiandeniz2018icc, hassanzadeh2016cache}. In \cite{hassanzadeh2016correlation}, correlation among an arbitrary number of files in a caching system is exploited by identifying the most representative files, which are then used as references for compressing the remaining correlated files. A two-file system is studied in detail in \cite{hassanzadeh2017rate}, where the files are initially compressed using Gray-Wyner source coding, and an optimal caching scheme is derived for the two-receiver network. This scheme is generalized to more files in \cite{hassanzadeh2017broadcast}, which is shown to be optimal for large cache sizes. Arbitrary numbers of users and files are considered in \cite{qiandeniz2018icc}, where each file is modeled as a collection of subfiles, each of which is shared by a different subset of files. In \cite{hassanzadehJSAC} a caching scheme is proposed, which exploits correlation by performing on-demand compression during delivery, resulting in a scheme that is robust to dynamic changes in a correlated library. Cache-aided content delivery is studied in \cite{qiandeniz2018tit} with a source coding perspective, where users request different quality lossy versions of the files in the library. More recently, \cite{MohammadDenizerasureTCom, amiri2017gaussian, shirin2016broadcastit, shirin2017broadcastit, Shengerasure2016}, and others have relaxed the assumption of a shared error-free bit-pipe from the server to the users by replacing it with noisy BCs in the traditional setting of uncorrelated files. In \cite{shirin2017broadcastit} the authors consider a total memory budget, and optimize the cache assignment with respect to the channel capacity for degraded BCs. In \cite{amiri2017gaussian} the benefits of caching and coded delivery are studied from the perspective of energy-efficiency in Gaussian BCs.


In this paper, we consider the degraded Gaussian BC model studied in \cite{amiri2017gaussian}, but we assume that the files in the library can be arbitrarily correlated as modeled in \cite{qiandeniz2018icc}. We evaluate the performance in terms of the minimum transmission power required to satisfy any demand combination, called the \textit{peak power} in \cite{amiri2017gaussian} as it represents the power required to satisfy the worst demand combination. We derive a lower bound on the minimum required transmission power assuming uncoded cache placement, and an upper bound that employs superposition coding and power allocation when delivering coded contents over the BC. Through numerical simulations, we show that the proposed correlation-aware joint caching and channel coding scheme reduces the transmission power significantly compared to schemes that are correlation-ignorant.

\textit{Notations:} $\mathbbm{R}$ is the set of all the real values, while $\mathbbm{R}^+$ denotes the set of all the positive real values. The set of integers $\left\{ i, ..., j \right\}$, where $i \le j$, is denoted by $\left[ i:j \right]$, and for $q\in \mathbbm{R}^+$, the set $[1: \lceil q\rceil]$ is denoted shortly by $[q]$. For sets $\mathcal{A}$ and $\mathcal{B}$, we define $\mathcal{A} \backslash \mathcal{B}\triangleq\{x: x \in \mathcal{A}, x\notin \mathcal{B}\}$, and $\left| \mathcal{A} \right|$ denotes the cardinality of $\mathcal{A}$. $\binom{j}{i}$ represents the binomial coefficient if $j\geq i$; otherwise, $\binom{j}{i}=0$. For event $E$, $\mathbbm{1}\{E\}=1$ if $E$ is true; and $\mathbbm{1}\{E\}=0$, otherwise.


\section{System Model}\label{sys}
Consider a wireless server that holds a database of $N$ correlated files, denoted by $\mathbf{W}=(W_1, ..., W_N)$, which have been jointly compressed into independent subfiles. File $W_i$, $i\in [N]$, consists of $2^{N-1}$ independent subfiles, i.e.,
\small
\begin{equation}
W_i=\{\overline{W}_{\mathcal{S}}: \mathcal{S}\subseteq[N],\;i \in \mathcal{S}\},  
\end{equation}
\normalsize
where $\overline{W}_{\mathcal{S}}$ denotes the subfile that is shared exclusively by the files $\{W_i: i \in \mathcal{S}\}$. The subfiles are arranged into $N$ {\em sublibraries}, $L_1,\dots,L_N$, such that $L_\ell$ contains all the subfiles that are shared exclusively by a set of $l$ files, i.e., 
\small
\begin{equation}
L_\ell=\{\overline{W}_{\mathcal{S}}:\mathcal{S}\subseteq [N],\; |\mathcal{S}|=\ell\}.
\end{equation}
\normalsize
Motivated by Gray-Wyner compression, we assume that all the subfiles in a sublibrary have the same length, and subfile $\overline{W}_{\mathcal{S}}\in L_\ell$ is distributed uniformly over the set $[2^{  nR_\ell }]$, where $R_\ell$ is the rate of subfile $\overline{W}_{\mathcal{S}}$ in bits per channel use, and $n$ denotes the transmission blocklength, referring to $n$ uses of the BC. We define $\mathbf{R} \triangleq (R_1, \ldots, R_N)$. All the files are of the same normalized rate of $R$ bits per channel use, where
\small
\begin{equation}
R=\sum\limits_{\ell=1}^{N} \binom{N-1}{\ell-1}R_\ell.
\end{equation}
\normalsize
 

Each user is equipped with a cache of size $nM$ bits, where $M$ is called the \textit{normalized cache capacity}. Communication takes place in two phases. During the first phase, referred to as the {\em placement phase}, the user caches are filled by the server without the knowledge of particular demands. This phase happens during a period of low traffic, which can be considered noiseless. In this paper, we are restricted to {\em uncoded cache placement phase}, i.e, users only caches uncoded messages. We consider {\em centralized} caching; that is, the server has the knowledge of the active users in advance, allowing the cache placement to be conducted in a coordinated fashion. At the beginning of the second phase, referred to as the {\em delivery phase}, user $k\in[K]$ requests file $W_{d_k}$ from the library, with $d_k$ uniformly distributed over $[N]$. Let $\mathbf{d}\triangleq (d_1, ..., d_K)$ denote the demand vector. All the requests are satisfied through a Gaussian BC, characterized by a time-invariant channel vector ${\bf h} =(h_1 ,\dots,h_K)$ and additive white Gaussian noise, where $h_k$ denotes the real channel gain between the server and user $k$. The channel gains are fixed, and are known to all the parties. Without loss of generality, we assume $h_1^2 \leq h_2^2 \leq \cdots \leq h_K^2$, such that the users are ordered from the weakest to the strongest. The $i^\text{th}$ channel output, is given by
\begin{equation}\label{channel}
Y_{k,i}=h_k\, X_i+\sigma_{k,i}, 
\end{equation}
 for channel gain $h_k$, and noise $\sigma_{k,i}\sim \mathcal{N}(0,1)$ at user $k$ in the $i^\text{th}$ channel use, which is assumed to be independently and identically distributed across time and users. 

An $(n, \mathbf{R}, M, P)$ code for this system consists of:
\begin{itemize}
\item \textbf{$K$ caching functions} $f_{k}$, $k \in [K]$,
\begin{equation}
  f_{k}: [2^{nR}]^N\times \mathbbm{R}^{K} \rightarrow [2^{nM}],
\end{equation}
and user $k$'s cache content, denoted by $Z_k$, is given by  $Z_k=f_{k}(\mathbf{W},\mathbf{h})$. Let $\mathbf{Z}\triangleq (Z_1,\dots,Z_K)$
\item A \textbf{delivery function} $g$, 
\begin{equation}
g: [ 2^{nR}]^N \times [ 2^{nM}]^K\times  \mathbbm{R}^{K} \times [N]^K \rightarrow \mathbbm{R}^{n},
\end{equation}
which, for given cache contents $\mathbf Z$, channel gains $\mathbf{h}$, and demand vector $\bf d$, generates the channel input signal,  $X^n(\mathbf{W}, \mathbf{d})=g(\mathbf{W},\mathbf Z, \mathbf{h}, \mathbf{d})$, transmitted over the Gaussian BC in $n$ channel uses, with $X_i(\mathbf{W}, \mathbf{d})$ denoting the $i^\text{th}$ channel input. The channel input vector is generated such that its average power is not more than $P$, i.e., 
\begin{equation}
P(\mathbf{W}, \mathbf{d}) \triangleq \sum\limits_{i=1}^n X^2_i(\mathbf{W}, \mathbf{d})\leq P, \quad \forall\, {\bf d}\in [N]^K.
\end{equation}
\item \textbf{$K$ decoding functions} $\phi_k$, $k [K]$,
\begin{equation}
\phi_k: \mathbbm{R}^{n}   \times [2^{nM}]\times \mathbbm{R}^{K} \times [N]^K \rightarrow [2^{nR}],
\end{equation}
where $\widehat{W}_{d_k}=\phi_k(Y^n,Z_k,\mathbf{h},\mathbf{d})$, is the reconstruction of $W_{d_k}$, and $Y^n$ is the channel output at user $k$. 
\end{itemize}

\begin{Definition}
A cache capacity-power pair $(M, P)$ is \textit{achievable} for the system described above, if there exists a sequence of $(n, \mathbf{R}, M, P)$ codes such that 
\small
\begin{equation}
\lim_{n \rightarrow \infty} \mathbbm{P} \Bigg\{\bigcup\limits_{{\bf d}\in [N]^K } \bigcup\limits_{k =1}^K \Big\{\widehat{W}_{d_k}\neq W_{d_k}\Big\}\Bigg\}=0.
\end{equation}
\normalsize
\end{Definition}
For a given cache capacity $M$, our goal is to characterize the minimum power $P$ such that $(M, P)$ is achievable. The corresponding \textit{memory-power function} is defined as \small
\begin{equation}
    P^*(M)\triangleq \inf\{P: (M, P) \mbox{ is achievable}\}.
\end{equation}
\normalsize
We conclude the section with the following proposition, which is frequently referred to in the remainder of this paper.

 \begin{proposition}\label{prop:AWGN}
 \cite{bergmans1973degradedBC} 
 In a $K$-user degraded Gaussian BC with $h_1^2\leq h_2^2\leq\dots\leq h_K^2$, $K$ distinct messages with rates $\rho_1,\dots,\rho_K$, can be reliably transmitted to users $1,\dots,K$, respectively,  iff 
 \small
 \begin{equation}
\rho_k \leq C\Bigg(\frac{h_k^2P_k}{1+h_k^2\sum\limits_{j=k+1}^{K}P_j}\Bigg),~~~  k =1,\dots,K, \label{eq:codebook}
\end{equation}
\normalsize
 for some $P_1, ..., P_K$, which can be achieved by using superposition coding with Gaussian codewords of power $P_1,\dots,P_K$, where $C(x) \triangleq \frac{1}{2} \log_2(1+x)$. The minimum total transmit power for reliable communication is thus given by
\small
\begin{equation}
P=\sum\limits_{k=1}^K P_k \geq \sum\limits_{k=1}^{K}\left(\frac{2^{2\rho_k}-1}{h_k^2}\right) \prod\limits_{j=1}^{k-1}\frac{2^{2\rho_j}}{h_j^2}.\label{eq:power}
\end{equation}
\normalsize
\end{proposition}

\section{Main Results}
This section provides a lower and an upper bounds on the memory-power function, $P^*(M)$.

\subsection{Lower Bound on $P^*(M)$}
The following lower bound is stated here without a proof, which will be provided in a longer version of this paper. 
\begin{Theorem}\label{lowerbound}
For the caching problem described in Section~\ref{sys} with uncoded cache placement phase, the optimal memory-power function, $P^*(M)$, is lower bounded as
\small
\begin{align}
P^*(M)\geq P_{LB}(M)\triangleq\sum\limits_{k=1}^{\min\{N, K\}}\left(\frac{2^{2\tilde{\rho}_k}-1}{h_k^2}\right) \prod\limits_{j=1}^{k-1}\frac{2^{2\tilde{\rho}_j}}{h_j^2},
\end{align}
\normalsize
where
\small
\begin{align}\label{rstar}
\tilde{\rho}_k\triangleq \max\left\{\sum\limits_{\ell=0}^{N-k}\binom{N-k}{\ell}R_{\ell+1}-M, 0\right\},  \forall\, k \in[K].
\end{align}
\normalsize
\end{Theorem}

\subsection{Upper Bound on $P^*(M)$}
\begin{Theorem}\label{thm:ach power}
For the caching problem described in Section~\ref{sys}, the optimal memory-power function, $P^*(M)$, is upper bounded as
\small
\begin{align}
P^*(M)\leq  &\min\limits_{ {\boldsymbol \pi} = (\pi_1,\dots,\pi_N)} P_{UB}(M,  {\boldsymbol\pi}),\notag\\
&\qquad \mathrm{s.t.~}\quad   \sum\limits_{i=1}^N \pi_i\leq 1,\notag\\
&\qquad\qquad\quad  0\leq \pi_i \leq 1, ~~ i = 1,\dots,N,  \notag
\end{align}
\normalsize
where
\small
\begin{subequations}
\begin{align}
& P_{UB}(M,  {\boldsymbol\pi})\triangleq\sum\limits_{k=1}^{\min\{N, K\}}\left(\frac{2^{2\hat{\rho}_k}-1}{h_k^2}\right) \prod\limits_{j=1}^{k-1}{2^{2\hat{\rho}_j}}, \\
&\hat{\rho}_k\triangleq \sum\limits_{\ell=1}^N \sum\limits_{r=\max\{\ell-N+K,\,1\}}^{\min\{\ell, K\}  }\binom{N-K}{\ell-r}\binom{\min\{N, K\}-1}{r-1}\gamma_{k, \ell, r}\\
& {\gamma}_{k,\ell, r}\triangleq \notag\\
 &\begin{cases}
  \Big( \frac{\lfloor t_\ell\rfloor-t_\ell+1}{\lfloor t_\ell\rfloor-t_\ell} \lambda(\lfloor t_\ell\rfloor)   -\lambda(\lfloor t_\ell+1\rfloor )\Big) R_\ell,\, \mathrm{if}\; k\in[   \frac{\min\{N, K\}}{r} +1 ] \\
\qquad\qquad0, \qquad\qquad\qquad\qquad\qquad\qquad\quad \mathrm{otherwise}
\end{cases}\notag\\
& \lambda(t) \triangleq  \frac{\binom{K-k}{\lfloor t\rfloor}}{\binom{K}{\lfloor t \rfloor}}(\lfloor t \rfloor-t ),  \\
& t_{\ell} \triangleq \frac{K \pi_\ell M}{\binom{N}{\ell}R_\ell}.
\end{align}
\end{subequations}
\normalsize
\begin{proof}
The proof can be derived by characterizing the transmit power achieved by the coding scheme outlined in Section~\ref{sec:scheme}, and will be provided in the longer version.
\end{proof}
\end{Theorem}

\section{Cache-Aided Superposition Coding} \label{sec:scheme}
The upper bound in Theorem~\ref{thm:ach power} is achieved by a centralized caching and delivery scheme, which optimizes the cache allocation among different libraries and places uncoded messages into users' caches in a centralized manner. Superposition coding is employed to losslessly deliver coded messages over the Gaussian BC \cite{bergmans1973degradedBC}, where the coded messages are generated taking into account the correlation among the requested files as well as the channel gains. As in \cite{qiandeniz2018icc,hassanzadeh2017rate,hassanzadeh2017broadcast}, the scheme operates by treating the sublibraries independent of during the placement and delivery phases to determine the cache content and messages  targeted at each user, which are then jointly delivered over the BC. In the following, we first illustrate the main idea through a simple example, and then provide the general description of the coding  scheme.

\subsection{Motivating Example}\label{sec:example}
Consider a system with $K=3$ users having channel gains $h_1^2\leq h_2^2\leq h_3^2$, and a database of $N=3$ files with sublibraries: 
\begin{itemize}
\item $L_1 = \{\overline{W}_{\{1\}}, \overline{W}_{\{2\}},\overline{W}_{\{3\}}   \}$, each with rate $R_1$.
\item $L_2 = \{\overline{W}_{\{1,2\}}, \overline{W}_{\{2,3\}},\overline{W}_{\{1,3\}} \} $, each with rate $R_2$.
\item $L_3 = \{\overline{W}_{\{1,2,3\}}\}$, with rate $R_3$.
\end{itemize}
Each user has a cache with capacity $M = R_1+R_2+\frac{1}{3}R_3$. 

$\circ$ \textbf{Placement Phase:}
Since content placement is carried out independently across the sublibraries, for ease of exposition, we assume that the cache capacity is divided into three portions, such that the users allocate portions with normalized capacities $R_1$, $R_2$, and $\frac{1}{3}R_3$ for caching files from sublibraries $L_1$, $L_2$ and $L_3$, respectively. We remark that the cache capacity allocation in this example is not optimal, and the proposed scheme further optimizes the allocation as described in more detail in Subsection~\ref{sec:scheme general}. 
We use the prefetching policy proposed in~\cite{yu2017exact}, which divides the subfiles in sublibrary $L_\ell$ into three non-overlapping packets of size $\frac{1}{3}nR_\ell$ bits. Then, user $k$ caches 
\small
\begin{align}
Z_k=\Big\{\overline{W}_{\{1\}, \{k\}},& \overline{W}_{\{2\}, \{k\}}, \overline{W}_{\{3\}, \{k\}}, \overline{W}_{\{1, 2\}, \{k\}},\notag\\
 &\overline{W}_{\{2, 3\}, \{k\}}, \overline{W}_{\{1, 3\}, \{k\}}, \overline{W}_{\{1, 2, 3\}, \{k\}}\Big\}, \notag
\end{align}
\normalsize
 where $\overline{W}_{\mathcal{S}, \{k\}}$ denotes the $k^\text{th}$ packet of subfile $\overline{W}_{\mathcal{S}}$ cached at user $k\in [3]$.
 
 $\circ$ \textbf{Delivery Phase:}
 During the delivery phase, once the demand vector is revealed, the server computes the messages intended for each user, independently from each sublibrary, and then delivers them over the BC via superposition coding. Consider demand vector $\mathbf{d}=(1, 2, 3)$. User 1, who is the weakest user, requires all parts of subfiles $\{\overline{W}_{\{1\}},\overline{W}_{\{1,2\}},\overline{W}_{\{1,3\}},\overline{W}_{\{1,2,3\}}\}$ that are missing from its cache, in order to reconstruct $W_1$. User 2 requires the four subfiles corresponding to file $W_2$, but having a better channel compared to user 1, through successive cancellation, it can also decode the messages targeted at user 1. Similarly, user 3 can decode both messages indented for the weaker users. User messages from each sublibrary are determined as follows.
 
 \begin{itemize}
 \item Sublibrary $L_1$: Based on the demand, all subfiles in $L_1$ are required by the users. User 1 needs to receive $\overline{W}_{\{1\}, \{2\}}$ and $\overline{W}_{\{1\}, \{3\}}$, whose targeted message, denoted by $V^1_{\mathbf{d}}(L_1)$, is generated as follows:
 \small
 \begin{align}
 V^1_{\mathbf{d}}(L_1)= \{&
 \overline{W}_{\{1\}, \{2\}}\oplus\overline{W}_{\{2\}, \{1\}}, \overline{W}_{\{1\}, \{3\}}\oplus\overline{W}_{\{3\}, \{1\}}\}.
 \label{eq:V1 L1}
 \end{align}
 \normalsize
 Since user 2 is able to decode its required packet $\overline{W}_{\{2\}, \{1\}}$ from message $V^1_{\mathbf{d}}(L_1)$, it only needs $\overline{W}_{\{2\}, \{3\}}$, which is recovered through the message
 \small
 \begin{align}
 V^2_{\mathbf{d}}(L_1)=\left\{\overline{W}_{\{2\}, \{3\}}\oplus\overline{W}_{\{3\}, \{2\}} \right\}.\label{eq:V2 L1}
 \end{align}
 \normalsize
 User 3 can decode its missing packets from $V^1_{\mathbf{d}}(L_1)$ and $V^2_{\mathbf{d}}(L_1)$, and therefore, $V^3_{\mathbf{d}}(L_1)=\emptyset$. 
 \item Sublibrary $L_2$: Each user requires two subfiles from $L_2$, which can be considered as two separate demands. One possible partitioning could be $\mathfrak{S}_1 = (\{1,2\}, \{1,2\}, \{1,3\})$ and $\mathfrak{S}_2=(\{1,3\}, \{2,3\}, \{2,3\})$, where $\mathfrak{S}_1$ corresponds to users 1, 2 and 3 requesting subfiles $\overline{W}_{\{1,2\}}$, $\overline{W}_{\{1,2\}}$ and $\overline{W}_{\{1,3\}}$, respectively. 
 Then $V^k_{\mathbf{d}}(L_2)=\{v^k_1 , v^k_2$\}, where $v^k_i$ is user $k$'s message corresponding to demand $\mathfrak{S}_i$. Then, for $\mathfrak{S}_1$
 \small
  \begin{align}
 &v^1_1= 
 \{\overline{W}_{\{1,2\}, \{2\}}\oplus\overline{W}_{\{1,2\}, \{1\}},
 \overline{W}_{\{1,2\}, \{3\}}\oplus\overline{W}_{\{1,3\}, \{1\}}\},\label{eq:V1 L2 a}\\
 &v^2_{1}=\left\{\overline{W}_{\{1,3\}, \{2\}}\oplus\overline{W}_{\{1,2\}, \{3\}} \right\},\label{eq:V2 L2 a}\\
 & v^3_{1}=\emptyset,\label{eq:V2 L2 a}
 \end{align}
 \normalsize
 and for $\mathfrak{S}_2$
 \small
 \begin{align}
  &v^1_2=
 \{\overline{W}_{\{1,3\}, \{2\}}\oplus\overline{W}_{\{2,3\}, \{1\}},
 \overline{W}_{\{1,3\}, \{3\}}\oplus\overline{W}_{\{2,3\}, \{1\}}\},\label{eq:V1 L2 b}\\
 &v^2_{2}=\left\{\overline{W}_{\{2,3\}, \{2\}}\oplus\overline{W}_{\{2,3\}, \{3\}} \right\},\label{eq:V2 L2 b}\\
 & v^3_{2}=\emptyset.\label{2eq: V2 L2 b}
 \end{align}
 \normalsize 
 \item Sublibrary $L_3$: All users require $\overline{W}_{\{1,2,3\}}$, and therefore
 \small
  \begin{align}
 &V^1_{\mathbf{d}}(L_3)=
  \{\overline{W}_{\{1,2,3\}, \{2\}}\oplus\overline{W}_{\{1,2,3\}, \{1\}}, \notag\\
 &\qquad\qquad\qquad\qquad
 \overline{W}_{\{1,2,3\}, \{3\}}\oplus\overline{W}_{\{1,2,3\}, \{1\}} \},\label{eq:V1 L3}\\
 &V^2_{\mathbf{d}}(L_2)=  V^3_{\mathbf{d}}(L_2)=\emptyset. \label{eq:V2 L3}
 \end{align}
 \normalsize
 \end{itemize}
 The messages in \eqref{eq:V1 L1}, \eqref{eq:V1 L2 a}, \eqref{eq:V1 L2 b} and \eqref{eq:V1 L3} constitute all the messages targeted at user 1, with total rate $\rho_1=2(R_1+2R_2+R_3)$. Messages \eqref{eq:V2 L1}, \eqref{eq:V2 L2 a} and \eqref{eq:V2 L2 b} are targeted at user 2 with total rate $\rho_2=R_1+2R_2$, and finally, user 3 can successfully recover its requested file from the messages intended for users 1 and 2, i.e., $\rho_3=0$. Based on Proposition~\ref{prop:AWGN}, the target rates can be delivered to the users with superposition coding of Gaussian codewords  satisfying \eqref{eq:codebook}, with a minimum power value given in \eqref{eq:power}.

\subsection{Proposed Scheme}\label{sec:scheme general}
 This section presents the description of the proposed centralized caching and delivery scheme that generalizes the example in Subsection~\ref{sec:example}, and achieves the transmission power stated in Theorem~\ref{thm:ach power}. Similarly to the schemes in \cite{qiandeniz2018icc,hassanzadeh2017rate,hassanzadeh2017broadcast}, the proposed scheme treats the sublibraries independently: 1) the cache capacity is optimally divided among the $N$ sublibraries, and user caches are filled in a centralized fashion, 2) for each demand realization, the server identifies the messages that need to be delivered to each user, independently across sublibraries, using a modified version of the scheme proposed in \cite{qiandeniz2018icc}, and 3) the server employs superposition coding to reliably communicate coded messages over the Gaussian BC.

\subsubsection{\bf Placement Phase}\label{placementphase}
As in \cite{qiandeniz2018icc,hassanzadeh2017rate,hassanzadeh2017broadcast}, the contents placed in each user's cache are identified separately for the different sublibraries, each with a different level of commonness. To this end, a fraction of the available cache capacity is allocated to each sublibrary. Let $\boldsymbol \pi = (\pi_1,\dots,\pi_N)$ denote the cache allocation vector, where $\pi_\ell\in[0,1]$ denotes the fraction of $M$ allocated to sublibrary $L_\ell$, with $\sum_{\ell=1}^N \pi_\ell=1$. We will later optimize $\boldsymbol \pi$ to minimize the required total transmit power. For a given $\boldsymbol \pi$, placement for sublibrary $L_\ell$ is carried out using the prefetching scheme proposed in~\cite{yu2017exact} as follows. Let
\begin{equation}\label{tl}
t_\ell\triangleq \frac{K\pi_\ell M}{\binom{N}{\ell}R_\ell}, ~  t_\ell\in[0,K],
\end{equation}
which, differently from~\cite{yu2017exact}, is not necessarily an integer. We address this by memory-sharing among the neighboring integer points, $t_\ell^A \triangleq \lfloor t_\ell \rfloor$ and $t_\ell^B \triangleq\lfloor t_\ell \rfloor+1$, and divide each subfile $\overline{W}_{\mathcal{S}}\in L_\ell$ into two non-overlapping parts. More specifically, $\overline{W}_{\mathcal{S}}=(\overline{W}^A_{\mathcal{S}}, \overline{W}^B_{\mathcal{S}})$, where  $\overline{W}^A_{\mathcal{S}}$ is at rate $(t_\ell^B-t_\ell)R_\ell$, while $\overline{W}^B_{\mathcal{S}}$ is at rate $(t_\ell- t_\ell^A)R_\ell$. The prefetching policy of \cite{yu2017exact} is implemented separately for $\{\overline{W}^A_{\mathcal{S}}: \mathcal{S}\in L_\ell\}$ and $\{\overline{W}^B_{\mathcal{S}}: \mathcal{S}\in L_\ell\}$. Each part $\overline{W}^A_{\mathcal{S}}$ is split into $\binom{K}{t_l^A}$ non-overlapping equal-length packets, with size $n( t_\ell^B-t_\ell)/\binom{K}{ t_\ell^A} R_\ell$ bits. These packets are assigned to sets $\mathcal A \subseteq [K]$ with size $t_\ell^A$. We denote the packet assigned to $\mathcal A$ by $\overline{W}^A_{\mathcal{S}, \mathcal{A}}$; therefore,
\small
\begin{equation}
\overline{W}^A_{\mathcal{S}}= \{   \overline{W}^A_{\mathcal{S}, \mathcal{A}} :\mathcal{A}\subseteq [K] ,\, |\mathcal{A}|=  t_\ell^A \}. 
\end{equation}
\normalsize
   Similarly, each part $\overline{W}^B_{\mathcal{S}}$ is split into $\binom{K}{t_l^B}$ non-overlapping equal-length packets, which are labeled as 
 \small
\begin{equation}
\overline{W}^B_{\mathcal{S}}= \{   \overline{W}^B_{\mathcal{S}, \mathcal{B}} :\mathcal{B}\subseteq [K] ,\, |\mathcal{B}|=  t_\ell^B \}.
\end{equation}
\normalsize

Given this packetization, user $k$ caches packets $\overline{W}^A_{\mathcal{S}, \mathcal{A}}$   if $k \in \mathcal{A}$, and packets $\overline{W}^B_{\mathcal{S}, \mathcal{B}}$   if $k \in \mathcal{B}$. With this placement strategy,  for each subfile in sublibrary $L_\ell$,  $\binom{K-1}{ t_\ell^A-1}$ distinct packets from $\overline{W}^A_{\mathcal{S}}$, and $\binom{K-1}{t_\ell^B-1}$ distinct packets from  $\overline{W}^B_{\mathcal{S}}$, are placed in each user's cache, amounting to a total of $nt_\ell R_\ell/K$ bits, which satisfies the capacity constraint of $n\pi_\ell M$ bits. 

\subsubsection{\bf Delivery Phase}\label{deliveryphase}
As explained in \cite{qiandeniz2018icc, hassanzadeh2017rate, hassanzadeh2017broadcast}, delivering a file from a correlated library, which has been compressed into multiple subfiles, is a multiple-demand problem. For a given demand vector $\bf d$, user $k$ requires $ \binom{N-1}{\ell-1}$ subfiles from sublibrary $L_\ell$, in order to successfully reconstruct file $W_{d_k}$. Since the sublibraries are treated independently, message $V^k_{\mathbf{d}}$, targeted at user $k$, constitutes the messages computed from all the sublibraries, i.e.,
\small
\begin{equation}
V^k_{\mathbf{d}} =   \bigcup\limits_{\ell=1}^N V^k_{\mathbf{d}}(L_\ell),\label{eq:messages}
\end{equation}
\normalsize
where  $V^k_{\mathbf{d}}(L_\ell)$ denotes the set of messages from sublibrary $L_\ell$ targeted at user $k$. Messages $V^1_{\mathbf{d}}(L_\ell),\dots ,V^K_{\mathbf{d}}(L_\ell)$ are determined using Algorithm~1, which is based on \cite[Algorithms 1, 2]{qiandeniz2018icc}. The main idea is to treat subfiles $\{\overline{W}_{\mathcal S}: d_k \in \mathcal S\}$ that are not cached at user $k$, as different demands. The algorithm operates by partitioning all the requested subfiles from sublibrary $L_\ell$ into groups, such that each user requires at most one subfile in each group; thus, resulting in a single-demand problem. 

For sublibrary $L_\ell$, $V^1_{\mathbf{d}}(L_\ell),\dots, V^K_{\mathbf{d}}(L_\ell)$, are generated as follows:
  \begin{itemize}
      \item[$i)$] Grouping the requested subfiles:
      
      Let $\mathcal{D}\triangleq\{d_1, ..., d_K\}$ denote the set of distinct files requested in demand vector $\bf d$. The subfiles that need to be delivered to at least $\ell$ users, are given by: 
      \begin{equation}\label{eq:set for empty}
      \{\overline{W}_{\mathcal{S}}:\overline{W}_{\mathcal{S}}\in L_\ell,  \mathcal{S} \subseteq\mathcal{D}\}.
      \end{equation}
       Since each user can request multiple subfiles from \eqref{eq:set for empty}, they are grouped into multiple (possibly overlapping) sets with minimum cardinality, such that each group represents the demand set of a single-demand network with K users. The grouping process tries to minimize the number of distinct demands within each corresponding single-demand network. For sublibrary $L_\ell$, where each subfile is required by $\ell$ distinct demands, there are at most $\lceil |\mathcal{D}| /\ell\rceil+1$ subfiles in each group. Note that, the subfiles in \eqref{eq:set for empty} are not the only contents that need to be delivered from sublibrary $L_\ell$. Based on the demand vector, any subfile $\overline{W}_{\mathcal{S}}$ whose index $\mathcal{S}$ includes at least one of the indices in $\mathcal D$, i.e.,  $\mathcal{S}\cap\mathcal{D}\neq \emptyset$, is required for the lossless reconstruction of the corresponding requested file in $\mathcal{D}$. All such subfiles need to be identified, and grouped in a similar fashion. The subfiles in \eqref{eq:set for empty} correspond to $|\mathcal{S}\cap\mathcal{D}|=\ell$. For $r=1,\dots,\ell$, we define the requested subfiles  ${\mathcal W}_r$, as 
      \begin{equation}\label{eq:groups}
      {\mathcal W}_r \triangleq \{\overline{W}_{\mathcal{S}}: |S| = \ell,\; |\mathcal{S}\cap\mathcal{D}|=r \}.
      \end{equation}
      Then, each set ${\mathcal W}_r$ is grouped using the function GROUP in Algorithm~1, which assigns a demand vector $\mathfrak{S}=({\mathcal S}_1,\dots,{\mathcal S}_K)$ to each group, resulting in a single-demand network with $K$ users, where user $k$ requests subfile $\overline{W}_{\mathcal{S}_k}$.

      \item[$ii)$] Delivering the demands corresponding to each group:\\ The groups that have been formed above are treated independently in the delivery phase. More specifically, for a group with corresponding demand vector $\mathfrak{S}$, function SINGLE-DEMAND in Algorithm~1 identifies the messages $V^1,\dots,V^K$ that need to be transmitted so that all the users recover their requested subfiles in $\mathfrak{S}$. These messages are computed using the scheme in \cite{yu2017exact}, applied to a degraded BC. The channel is taken into account by selecting the {\em weakest} users with distinct demands as {\em leaders}, i.e., the demand of a leader is not requested by any of the weaker users, $ \{k: \mathcal{S}_{k}\notin \{\mathcal{S}_{1}, ..., \mathcal{S}_{k-1}\}\}$, and then greedily broadcasting XORed messages that benefit at least one leader. Note that choosing the weakest user, among users requiring the same subfile $\overline{W}_{\mathcal S}$, as the leader, allows all the stronger users to decode the subfile through superposition coding and successive cancellation decoding. As mentioned previously, the proposed scheme uses memory-sharing to cache and deliver the subfiles in $L_\ell$, for the two parts $\overline{W}_{\mathcal{S}}^A$ and $\overline{W}_{\mathcal{S}}^B$; and therefore, function SINGLE-DEMAND is executed for both parts. 
   \end{itemize}
   \begin{algorithm}[H]\label{message:1}
\caption{Generate messages $\{V^1_{\mathbf{d}}(L_\ell),\dots, V^K_{\mathbf{d}}(L_\ell)\}$}
\label{groupingscheme}
\begin{algorithmic}[1]
\Statex
\small
\State{$V^k_{\mathbf{d}}(L_\ell) \leftarrow  \emptyset$, $\forall k \in \{1,\dots,K\}$}
\For{$r \in \{1 ,\dots,\ell\}$}
\State{\small ${\mathcal W}_r= \{\overline{W}_{\mathcal{S}}: |S| = \ell,\; |\mathcal{S}\cap\mathcal{D}|=r \}$}
\State{$\mathfrak{S}_1,\dots, \mathfrak{S}_g$ $\leftarrow$ Group (${\mathcal W}_r$, $\mathcal{D}$, $\ell$, $r$)}  
\For{$i \in \{1 ,\dots,g\}$}
\State{$V^1_A,\dots, V^K_A$ $\leftarrow$ Single-Demand ($A$, ${\mathfrak{S}_i}$, $t_\ell^A$)}
\State{$V^1_B,\dots, V^K_B$ $\leftarrow$ Single-Demand ($B$, ${\mathfrak{S}_i}$, $t_\ell^B$)}
\State{$V^k_{\mathbf{d}}(L_\ell)\leftarrow V^k_{\mathbf{d}}(L_\ell) \cup \{ V^k_A,V^k_B\}$, $\forall k \in \{1 ,\dots,K\}$}
\EndFor
\EndFor
\end{algorithmic}
\end{algorithm}
  Message $V^k_{\mathbf{d}}(L_\ell)$ targeted at user $k$ is the union of all the messages for sublibrary $L_\ell$ computed for each group identified from the subfile sets $\{\mathcal W_1,\dots, \mathcal W_\ell\}$.The overall message for user $k$, $V^k_{\mathbf{d}}$, is thus obtained by \eqref{eq:messages}. For a given demand vector $\bf d$, messages $V_{\mathbf d}^1,\dots,V_{\mathbf d}^K$ can be reliably transmitted to users $1,\dots,K$, using a $K$-level Gaussian superposition codebook \cite{bergmans1973degradedBC}. The $k^\text{th}$-level codebook consists of $2^{n\rho_{k}}$ codewords, where $\rho_k$ is the total rate of the messages in $V^{k}_{\mathbf{d}}$. The total required transmit power is given by \eqref{eq:codebook} in Proposition \ref{prop:AWGN}.

\begin{algorithm}

\begin{algorithmic}[1]
\Function {Group }{ ${\mathcal W}$, $\mathcal{D}$, $\ell$, $r$}\\
\textbf{Output:} Group demands $\mathfrak{S}_1,\dots, \mathfrak{S}_g$
\small
\State{$\mathcal{F} \leftarrow \mathcal{D}$, $\overline{\mathcal{F}}\leftarrow \emptyset$, $\overline{\mathcal{S}}\leftarrow \emptyset$,
$g=0$}
\While{$\mathcal{W}\neq \emptyset$}
\While{$\mathcal{F}\neq \emptyset$}
\If{$|\mathcal{F}|\geq r$}
\If{$\overline{\mathcal{F}}=\emptyset$}
\State{ Randomly pick \footnotesize{$\overline{W}_{\mathcal{S}}\in \mathcal{W}$} such that \footnotesize{$\mathcal{S}\cap{\mathcal{D}} \subseteq\mathcal{F}$}}
\State{
$\mathcal{W}\leftarrow \mathcal{W}/\overline{W}_{\mathcal{S}},\quad
\mathcal{F} \leftarrow  \mathcal{F}\setminus \mathcal{S}$}
\For{$d_k \in \mathcal{S}\cap{\mathcal{D}}$}
\State{$\mathcal{S}_{k} \leftarrow  \mathcal{S}$}
\EndFor
\Else {\For{$d_k \in \overline{\mathcal{F}}$}
\State{$\mathcal{S}_{k} \leftarrow  \overline{\mathcal{S}}$}
\EndFor}
\State{\small
$\mathcal{F} \leftarrow \mathcal{F}\setminus \overline{\mathcal{F}},\quad \overline{\mathcal{S}} \leftarrow  \emptyset,\quad \overline{\mathcal{F}} \leftarrow \emptyset, $}
\EndIf
\Else\State{\small Randomly pick $\overline{W}_{\mathcal{S}}\in \mathcal{W}$ such that $\mathcal{F}\subseteq \mathcal{S}$}
\For{$d_k \in \mathcal{F}$}
\State{$\mathcal{S}_{k} \leftarrow  \mathcal{S}$}
\EndFor
\small
\State{$\mathcal{F} \leftarrow \emptyset,\quad
 \overline{\mathcal{S}} \leftarrow  \mathcal{S},\quad \overline{\mathcal{F}} \leftarrow  \mathcal{S}\setminus  {\mathcal{F}}$}
\normalsize
\EndIf
\EndWhile
\State{$g=g+1$}
\State{$\mathfrak{S}_g= (\mathcal{S}_{1}, \dots, \mathcal{S}_{K})$}
\EndWhile
\EndFunction
\end{algorithmic}
\end{algorithm}

\begin{algorithm}
\begin{algorithmic}[1]
\Function{Single-Demand }{$C$, ${\mathfrak{S}}$, $t$}\\
\textbf{Input:} $\mathfrak{S}=(\mathcal{S}_{1}, \dots, \mathcal{S}_{K})$, $C\equiv \{\overline{W}^C_{\mathcal S,\mathcal C}\}$\\
\textbf{Output:} Coded messages $V^1,\dots,V^K$
\small
\State{$\mathcal{K} \leftarrow \{k: \mathcal{S}_{k}\notin \{\mathcal{S}_{1}, ..., \mathcal{S}_{k-1}\}\}$}
\For{$k\in \{1,\dots,K\}$}
\For{{\footnotesize$\mathcal{U}\subseteq [k+1:K]: |\mathcal{U}|= t, \sum\limits_{j \in \mathcal{K}}\mathbbm{1}\{j \in \mathcal{U}\cup\{k\}\} \geq 1$}}
\State{  $
V^k\leftarrow V^k \bigcup\left(\bigoplus\limits_{j\in \mathcal{U}\cup\{k\}} \overline{W}^C_{\mathcal{S}_{j}, \mathcal{U}\cup\{k\}\setminus \{j\}}\right)$}
\EndFor
\EndFor
\EndFunction

\end{algorithmic}
\end{algorithm}

\section{Numerical results}\label{s:numerical}
We evaluate the performance of the scheme proposed in Subsection~\ref{sec:scheme general}, referred to as the {\em correlation-aware} scheme, by comparing its memory-power trade-off with the lower bound presented in Theorem~\ref{lowerbound}, as well as with the trade-off achieved by the scheme proposed in \cite{amiri2017gaussian}, which does not exploit the correlation among the files, referred to as the {\em correlation-ignorant} scheme. We consider a setting with $N=5$ files, $K=5$ users, file rate $R=1$, and cache capacity $M=0.5$. The channel gains are modeled as $1/h_k^2=2-0.2(k-1)$, for $k=1, ..., 5$. We denote by $\alpha_\ell$ the file-length fraction that belongs to sublibrary $L_\ell$, i.e.,
\small
\begin{equation}
\alpha_\ell=\binom{N-1}{\ell-1}\frac{R_{\ell}}{R}, \quad \sum\limits_{\ell=1}^N \alpha_\ell=1.\notag 
\end{equation}
\normalsize

\begin{figure}
\centering
\label{fig:1}
\includegraphics[width=0.83\linewidth]{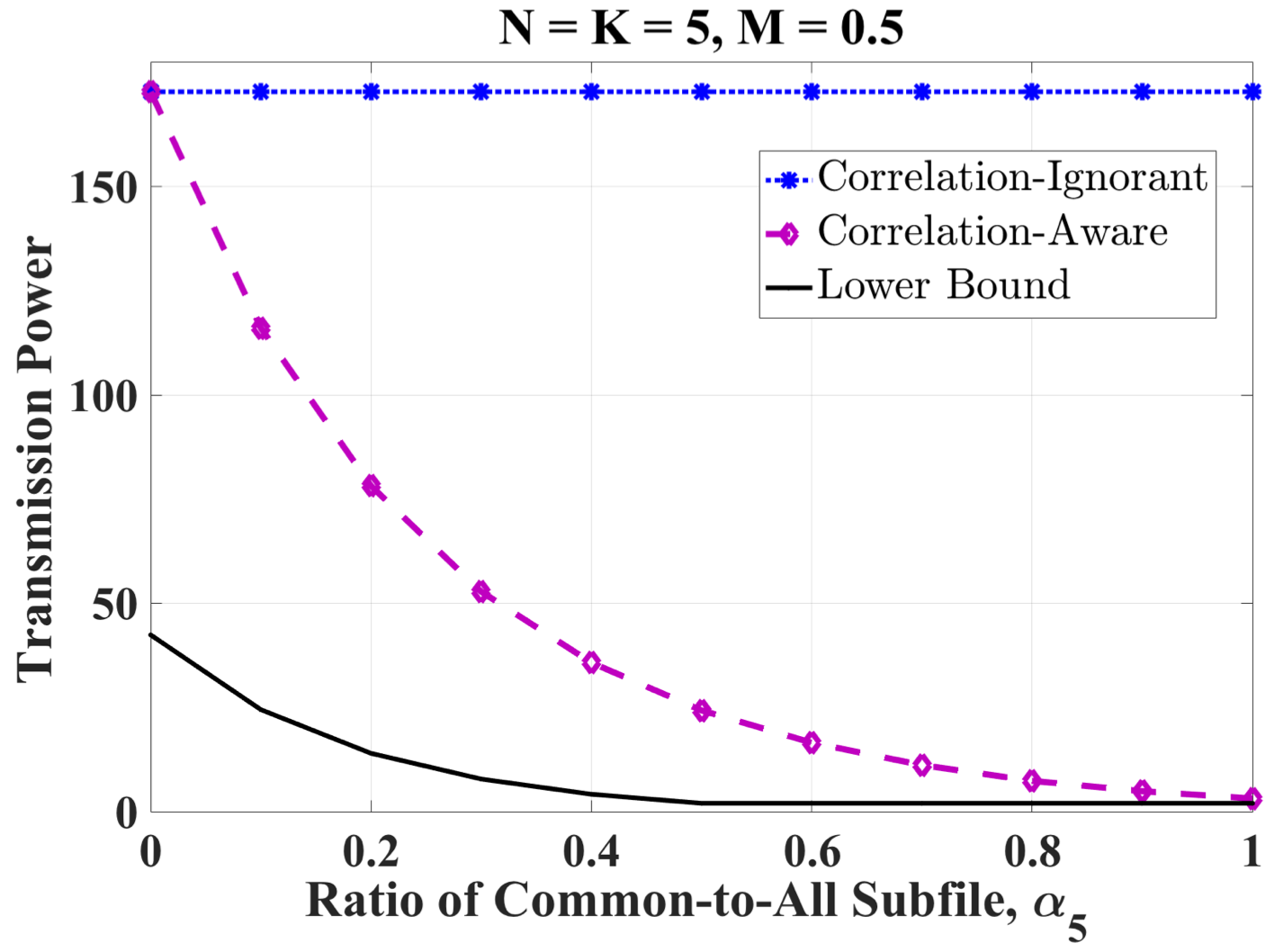}
\caption{Transmission power vs. common subfile fraction $\alpha_5$, when  the files are composed of private and common-to-all subfiles.}
\end{figure}
\begin{figure}

\label{fig:2}
\centering
\includegraphics[width=0.83\linewidth]{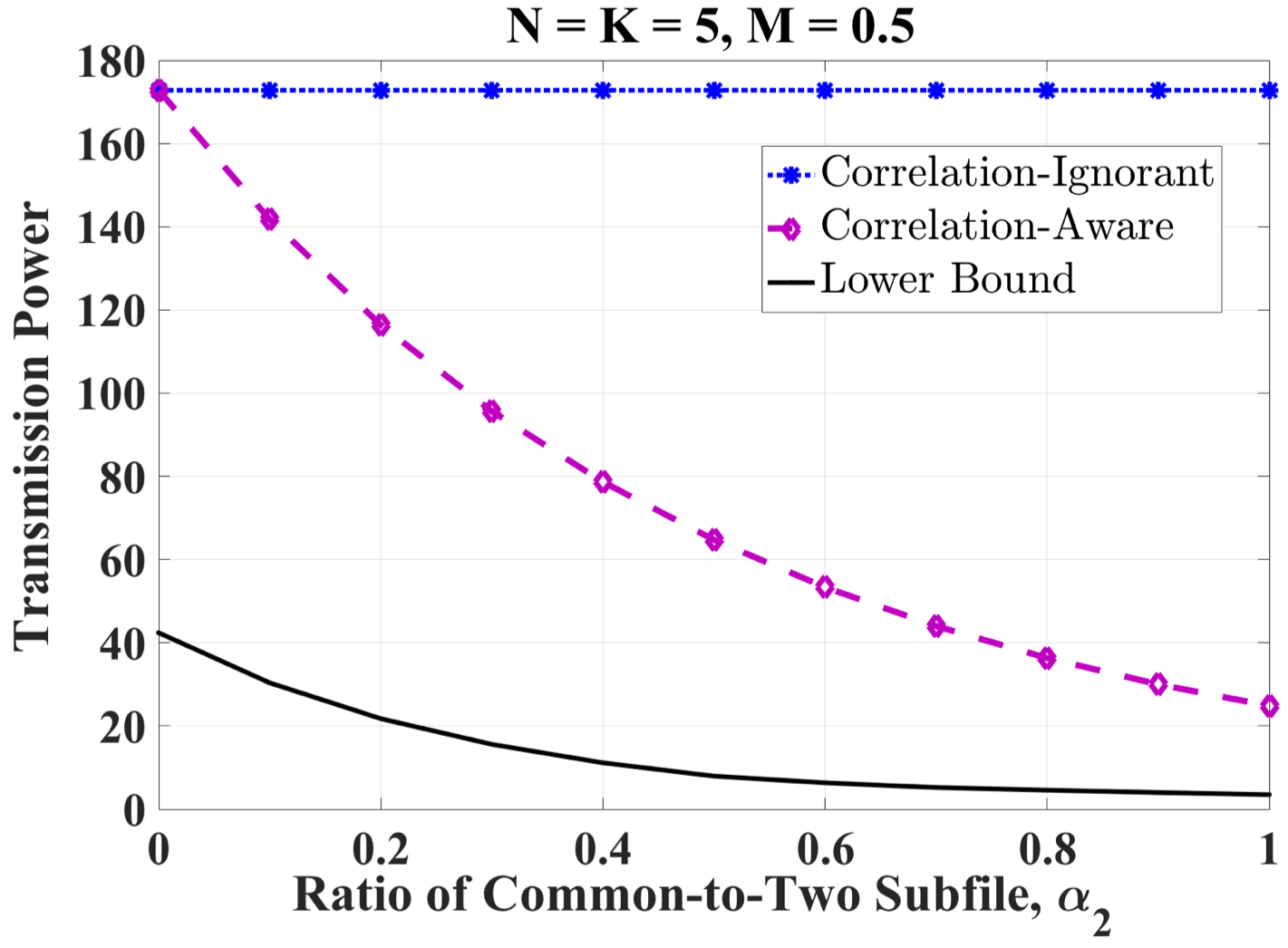}
\caption{Transmission power vs. common subfile fraction  $\alpha_2$, when  the files are composed of private and common-to-two subfiles.}
\end{figure}

Fig.~1 displays $p^*(M)$, minimum required transmit power for memory $M$, for a database with files composed of one {\em private} subfile, which is exclusive to that file, and a {\em common-to-all} subfile, which is shared among all files, i.e., $\alpha_1+\alpha_5=1$, $\alpha_2=\alpha_3=\alpha_4=0$. In Fig.~2, $p^*(M)$ is shown when the files, in addition to private subfiles, have pairwise correlations through {\em common-to-two} subfiles, that is $\alpha_1+\alpha_2=1$, $\alpha_3=\alpha_4=\alpha_5=0$. We plot the minimum transmit power as a function of the common parts of the files for both scenarios, i.e., with respect to $\alpha_5$ and $\alpha_2$. respectively. In both settings, the transmission power achieved by the correlation-aware scheme decreases remarkably, as the portion of common subfiles  increases, while the performance of the  correlation-ignorant scheme does not improve. It is observed that the transmission power drops at a higher rate in Fig.~1 compared to Fig.~2 for increasing ratio of common subfiles, in both the correlation-aware scheme and lower bound. This is due to the reduction in the amount of content that is sent over the Gaussian BC, as a result of exploiting the higher level of correlation. It is also observed that the gap between the transmit power upper and lower bounds is smaller in Fig.~1 compared to Fig.~2.

\section{Conclusions}
We have investigated caching and delivery of correlated contents over a cache-aided Gaussian BC. Correlation among files is captured by the component subfiles shared among different subsets of files. We have first presented a lower bound on the minimum transmission power which guarantees the reliable delivery of all possible demand combinations. A correlation- aware joint cache and channel coding scheme, based on superposition coding, is proposed, and the corresponding upper bound is compared numerically with the lower bound. 
Our numerical results show that significant energy gains can be obtained by exploiting the correlation among the files.

\bibliographystyle{IEEEtran}
\bibliography{report}
\end{document}